# Nano-patterned magnonic crystals based on ultrathin YIG films

M. Collet[1], M. Evelt[2], V. E. Demidov[2], J. L. Prieto[3], M. Muñoz[4], J. Ben Youssef[5], G. de Loubens[6], O. Klein[7], P. Bortolotti[1], V. Cros[1], A. Anane[1], and S. O. Demokritov[2,8]

[1]*Unité Mixte de Physique CNRS, Thales, Univ. Paris Sud, Université Paris-Saclay, 91767*

[2]*Institute for Applied Physics and Center for Nanotechnology, University of Muenster, 48149 Muenster, Germany*

[3]*Instituto de Sistemas Optoelectrónicos y Microtecnologa (UPM), Ciudad Universitaria, Madrid 28040, Spain*

[4]*IMM-Instituto de Microelectrónica de Madrid (CNM-CSIC), PTM, E-28760 Tres Cantos, Madrid, Spain*

[5]*Laboratoire de Magnétisme de Bretagne CNRS, Université de Bretagne Occidentale, 29285 Brest, France*

[6]*Service de Physique de l' État Condensé, CEA, CNRS, Université Paris-Saclay, CEA Saclay, 91191 Gif-sur-Yvette, France*

[7]*INAC-SPINTEC, CEA/CNRS and Univ. Grenoble Alpes, 38000 Grenoble, France*

[8]*M.N. Miheev Institute of Metal Physics of Ural Branch of Russian Academy of Sciences, Yekaterinburg 620041, Russia.*

We demonstrate a microscopic magnonic-crystal waveguide produced by nano-patterning of a 20 nm thick film of Yttrium Iron Garnet. By using the phase-resolved micro-focus Brillouin light scattering spectroscopy, we map the intensity and the phase of spin waves propagating in such a periodic magnetic structure. Based on these maps, we obtain the dispersion and the attenuation characteristics of spin waves providing detailed information about the physics of spin-wave propagation in the magnonic crystal. We show that, in contrast to the simplified physical picture, the maximum attenuation of spin waves is achieved close to the edge of the magnonic band gap, which is associated with non-trivial reflection characteristics of spin waves in non-uniform field potentials.

* Corresponding author, e-mail: demidov@uni-muenster.de

Magnonics[1-4] holds the promise of a paradigm change in the way signals are processed by coding the information into spin waves propagating in a magnetic medium. Nano-magnonics in particular aims at building a technological platform where lithographically defined devices are cascaded so to perform either digital computation, using spin wave interference as the computational scheme or on-chip analogue signal processing for radio-frequency applications. Relevant technological devices will require low loss materials to allow for long propagation distances of spin waves. Recent developments[6-9] in preparation of high-quality nanometer-thick films of the magnetic insulator Yttrium Iron Garnet (YIG) enabled significant advancements in the research field of magnonics. The unmatched small magnetic damping in this material allowed implementation of microscopic waveguiding structures exhibiting at least an order of magnitude larger propagation length of spin waves[10-14] in comparison with those based on metallic ferromagnets[3,15]. Together with the recently demonstrated highly efficient controllability of spin waves in ultrathin YIG by the spin transfer torque[16-18], the large spin-wave propagation length makes YIG one of the most promising materials for advanced nano-magnonic applications.

Similarly to photonic crystals in optics[19], one of the important functional elements of magnonic circuits are the artificial magnonic crystals[2,20-24] – spin-wave propagation media, where the characteristics of waves are tailored by using spatially periodic modulation of magnetic properties. The most efficient magnonic crystals demonstrated up to now have been based on micrometer-thick YIG films (see recent reviews [22,24] and references therein), where spin waves can propagate up to the distances of several millimetres much larger than the characteristic spatial scale of the periodicity typically equal to hundreds of micrometers. For a long time, YIG-based magnonic crystals could not be downscaled to the nano-range dimensions because of the difficulties in preparation of high-quality nanometer-thick YIG films. Therefore, the research in the area of nano-scale magnonic crystals mostly concentrated on systems based

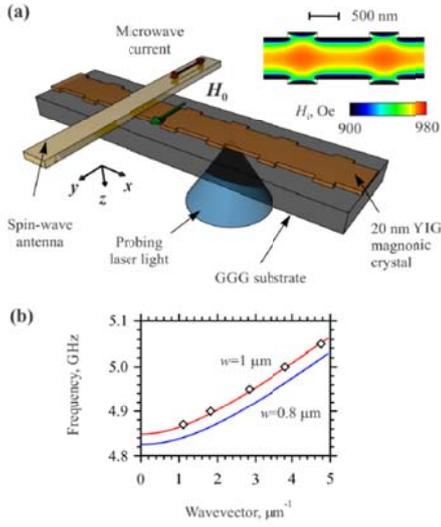

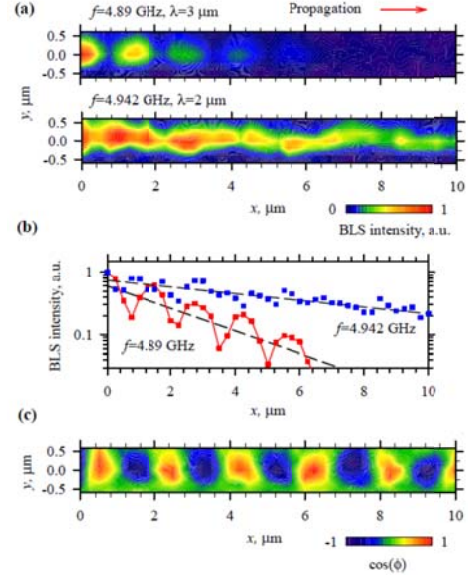

**Fig. 1** (a) Schematic of the experiment. Inset shows the distribution of the internal static magnetic field in the magnonic-crystal waveguide calculated by using the MuMax3 software[27]. (b) Solid curves – calculated dispersion curves for the 0.8 and 1 μm wide YIG stripes constituting the magnonic crystal. Symbols – experimentally determined dispersion for a regular, straight 1 μm wide reference waveguide fabricated from the same YIG film.

**Fig. 2** (a) Representative examples of two-dimensional spin-wave intensity maps recorded for spin-wave frequencies *f*=4.89 and 4.942 GHz corresponding to the wavelength λ=3 and 2 μm, respectively. (b) Propagation-coordinate dependence of the spin-wave intensity integrated across the width of the two-dimensional maps in the log-linear scale. Symbols – experimental data. Dashed lines – fit of the data by the exponential function. (c) Two-dimensional spin-wave phase map recorded for the frequency *f*=4.942 GHz.

on metallic magnetic films with the thicknesses of several tens of nanometers[2,20,21,23], where the propagation length of s pin waves does not exceed a few micrometers[3,15] making magnonic-crystal systems less efficient.

In this Letter, we demonstrate a magnonic crystal based on a 20 nm thick YIG film. The crystal is implemented in a form of a microscopic waveguide, whose width is periodically varied between 1 and 0.8 μm with the spatial period of 1.5 μm. We study the propagation characteristics of spin waves in this system by using the micro-focus Brillouin light scattering (BLS) spectroscopy, which allows space-resolved measurements of the intensity and the phase of propagating spin waves providing direct information about their attenuation and the wavelength. Based on these measurements, we reconstruct the dispersion curves of spin waves in a wide range of frequencies. We also directly determine the frequency interval of the magnonic band gap caused by the spatial periodicity and study the variation of the spin-wave attenuation across the gap. Our results show a non-trivial frequency dependence of the spin-wave decay constant within the band gap with the maximum attenuation corresponding to its upper edge. We associate these behaviors with the peculiarities of the spin-wave reflection, which is generally more complex than the reflection of light in photonic systems due to the influence of the non-local magnetic dipole interaction.

Figure 1(a) shows the schematic of the sample and the experimental set-up. The studied system is a width-modulated magnonic-crystal waveguide[25,26], patterned by e-beam lithography from a 20 nm thick YIG film grown by the pulsed laser deposition on gadolinium gallium garnet (GGG) (111) substrate[8]. Broadband ferromagnetic-resonance (FMR) measurements performed on extended films yield a Gilbert damping of $\alpha$=3.4 $10^{-4}$ with an extrinsic linewidth $\Delta H_0$ of 2 Oe and the effective magnetization $4\pi M_0$=2150 G. The width of the magnonic waveguide is periodically varied between 1 and 0.8 $\mu$m. The length of the narrow segments is 1 $\mu$m and that of the wide segments is 0.5 $\mu$m yielding in total the period of the modulation $a$=1.5 $\mu$m. The static magnetic field $H_0$ = 1000 Oe is applied perpendicular to the axis of the waveguide defining the propagation configuration of the so-called Damon–Eshbach spin waves. Due to the demagnetization effects, the modulation of the width leads to the periodic spatial modulation of the internal static magnetic field in the waveguide (inset in Fig. 1(a)). Although this modulation is relatively weak, it causes a noticeable modification of the spin-wave dispersion between the wide and narrow segments. To illustrate this fact, we show in Fig. 1(b) the dispersion curves for the 0.8 and 1 $\mu$m wide YIG stripes constituting the magnonic crystal calculated using the approach developed in Ref. 28 taking into account the demagnetizing fields. The validity of calculations is confirmed by the good agreement between the calculated curve and that obtained experimentally for a straight 1 µm wide YIG reference waveguide (symbols in Fig. 1(b)).

Spin waves in the magnonic waveguide are excited by using a 150 nm thick and 600 nm wide Au inductive antenna[29]. The space-resolved detection of the spin waves is performed using the micro-focus BLS technique[3]. The probing laser light with the wavelength of 473 nm and the power of 25 µW is focused through the sample substrate onto the surface of the YIG film into a diffraction-limited spot (Fig. 1(a)). The light scattered from spin waves is analysed by a high-contrast Fabry-Perot interferometer. The resulting signal – the BLS intensity – is proportional to the intensity of spin waves at the location of the probing spot. To additionally detect the phase of spin waves, we analyse the phase of the scattered light by using its interference with a reference light of the same frequency[30,31].

Figure 2(a) shows representative examples of spin-wave intensity maps recorded by rastering the probing laser spot over the waveguide for two different spin-wave frequencies corresponding to the wavelengths $\lambda$ = 3 and 2 µm. The maps clearly show that, due to the Bragg reflection, the spin wave with $\lambda$ = 3 µm = $2a$ forms a well-pronounced standing wave exhibiting a very fast spatial decay, as expected for waves with the wavevector $k' = k_B = \pi/a$ (Brillouin wavevector) propagating in a periodic potential with the period $a$. In contrast, the spin wave with $\lambda$ = 2 µm exhibits a much larger decay length. By plotting the propagation-coordinate dependence of the BLS intensity (Fig. 2(b)) and fitting the data by the exponential function (note log-linear scale), we determine the spin-wave decay constant (imaginary part of the wavevector) $k''$. The real part of the wavevector, $k'$, is found from the Fourier analysis of the phase maps (see Fig. 2(c)) recorded simultaneously with the intensity maps by using the phase-resolved BLS technique[30,31]. The phase maps reflect the spatial dependence of $\cos(\phi)$, where $\phi$ is the phase accumulated by the spin wave during its propagation from the antenna to the detection point.

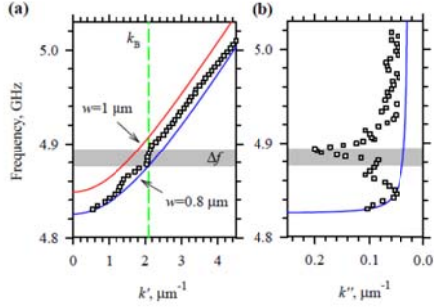

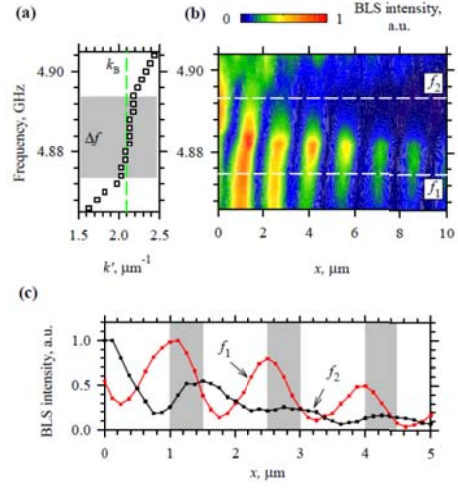

**Fig. 3** (a) Symbols – experimentally determined dispersion curve for spin waves in the magnonic-crystal waveguide. Solid curves – calculated dispersion curves for the 0.8 and 1 μm wide YIG stripes constituting the magnonic crystal. Vertical dashed line marks the Brillouin wavevector $k_B$. Shaded area marks the frequency region $\Delta f$ of the magnonic band gap. (b) Symbols – measured frequency dependence of the imaginary part of the wavevector $k''$ (decay constant). Solid curve – reference dependence $k''(f)$ calculated for 0.8 μm wide straight waveguide.

**Fig. 4** (a) Dispersion curve for spin waves at frequencies of the band gap determined from high-frequency-resolution BLS measurements. Vertical dashed line marks the Brillouin wavevector $k_B$. Shaded area marks the frequency region $\Delta f$ of the magnonic band gap. (b) Color-coded plot of the spin-wave intensity in the $x$-frequency coordinates. Dashed horizontal lines mark the edges of the band gap. (c) $x$-profiles of standing spin waves at frequencies $f_1$ and $f_2$ corresponding to the edges of the band gap. Shaded areas mark the positions of the wide segments of the magnonic-crystal waveguide.

Therefore, the spatial period of the phase map is equal to the wavelength of the spin wave at the given excitation frequency.

By performing the above-described measurements at different excitation frequencies $f$, we directly determine the frequency dependence of the real and the imaginary parts of the wavevector $k'(f)$ and $k''(f)$. As expected, the experimentally found dispersion curve $k'(f)$ (symbols in Fig. 3(a)) is located between the two curves calculated for the straight waveguides with the width of 1 (red curve) and 0.8 μm (blue curve). In the vicinity of the Brillouin wavevector $k_B = \pi/a = 2.1$ μm$^{-1}$, the slope of the measured dispersion curve exhibits an abrupt increase indicating the formation of the magnonic band gap. Note, that, in contrast to idealized lossless periodic systems, where there are no allowed spectral states within the band gap, in periodic systems with finite losses, the wave propagation is allowed within the band gap frequency range. In the latter case (see, e.g., 32, 33), the periodicity reveals itself in a degeneracy of the real part of the wavevector $k´ \approx k_B$ within the band gap, consistent with the data of Fig. 3(a). Additionally, the imaginary part of the wavevector $k´´$ is expected to abruptly increase in the band gap frequency range. The latter fact is illustrated in Fig. 3(b), which shows the measured dependence $k´´(f)$ (symbols) together with the reference dependence (solid curve) calculated for the straight waveguide using the Gilbert damping parameter $\alpha = 3.4 \times 10^{-4}$. Indeed, the data of Fig. 3(b) confirm a strong increase of $k´´$. However, contrarily to the simplified physical picture, the maximum of $k´´$ is not located at the center of the band gap, but is clearly shifted toward its upper frequency edge. The change in $k´´$ translates as almost a factor 4 decrease in the propagation length on a frequency span of about 15 MHz.

To clarify the observed behaviours, we perform high-frequency-resolution BLS measurements at frequencies of the magnonic band gap. The obtained dispersion curve shown in Fig. 4(a) allows precise identification of the band gap edges, where the $f(k´)$ dependence exhibits pronounced kinks. To analyse the peculiarities of the spin-wave propagation, we plot in Fig. 4(b) the color-coded spin-wave intensity versus the frequency and the propagation coordinate $x$. In this figure, one can see the modification of the $x$-profiles and of the spatial attenuation of standing spin waves with the variation of their frequency. In agreement with the data discussed above, the data of Fig. 4(b) show the strongest spin-wave decay at the upper frequency edge of the bandgap $f_2$, while the wave at the lower edge $f_1$ propagates to noticeably larger distances. From Fig. 4(b), one also sees that the positons of the maxima and minima of the standing wave change in space, when the frequency varies between $f_1$ and $f_2$. This fact is further illustrated in Fig. 4(c) showing the standing-wave profiles recorded at $f_1$ and $f_2$. As seen from these data, at the frequency $f_1$, the maxima of the standing wave are aligned with the left edges of the wide segments of the magnonic-crystal waveguide (shown by shaded areas). When the frequency is increased toward the upper edge $f_2$, the standing wave shifts by 0.5 μm and its maxima align with the right edges of the wide segments. This shift is accompanied by a strong increase in the spatial decay.

We would like to emphasize, that the spatial shift of the standing wave is also known for photonic crystals in optics[19]. However, in optical systems, it is not accompanied by the variation of the imaginary part of the wavevector, which typically maximizes at the frequency corresponding to the center of the band gap. The peculiar behavior observed for the magnonic crystal can be associated with the difference in the magnitude of the reflection coefficient for spin-wave reflection from the left and the right edges of the wide segments of the modulated-width waveguide. Indeed, in optical systems, the magnitude of the reflection coefficient is simply determined by the ratio between the refraction indexes of the layers constituting the periodic system. In contrast, the reflection characteristics of spin waves in non-uniform static field potentials differ strongly depending on whether the wave enters the region of larger or smaller fields[34]. This difference can be responsible for the asymmetric increase of the spatial decay at the upper edge of the magnonic band gap.

In conclusion, we have demonstrated a microscopic magnonic crystal based on ultra-thin YIG films suitable for implementation of nano-scale magnonic circuits with small dissipation losses. Our results clearly indicate a formation of the band gap in the studied system and show that behaviors of magnonic systems can differ significantly from those known in optics. Our observations should stimulate further experimental and theoretical research in the field of magnonic crystals and bring it closer to the real-world applications.

This work was supported in part by the Deutsche Forschungsgemeinschaft, the program Megagrant № 14.Z50.31.0025 of the Russian Ministry of Education and Science. M.C.

acknowledges DGA for financial support. We acknowledge E. Jacquet, R. Lebourgeois, R. Bernard,  A. H. Molpeceres and S. Xavier for their contribution to sample preparation.

# REFERENCES

1. V. V. Kruglyak, S. O. Demokritov, and D. Grundler, J. Phys. D Appl. Phys. **43**, 264001 (2010).

2. B. Lenk, H. Ulrichs, F. Garbs, and M. Münzenberg, Phys. Rep. **507**, 107–136 (2011).

3. V. E. Demidov and S. O. Demokritov, IEEE Trans. Mag. **51**, 0800215 (2015).

4. A.V. Chumak, V.I. Vasyuchka, A.A. Serga, and B. Hillebrands, Nature Phys. **11**, 453-461 (2015).

5. B. Divinskiy, V. E. Demidov, S. O. Demokritov, A. B. Rinkevich, and S. Urazhdin, Appl. Phys. Lett. **109**, 252401 (2016).

6. Y. Sun, Y.-Y. Song, H. Chang, M. Kabatek, M. Jantz, W. Schneider, M. Wu, H. Schultheiss, and A. Hoffmann, Appl. Phys. Lett. **101**, 152405 (2012).

7. C. Hahn, G. de Loubens, O. Klein, M. Viret, V. V. Naletov, J. Ben Youssef, Phys. Rev. B **87**, 174417 (2013).

8. O. d'Allivy Kelly, A. Anane, R. Bernard, J. Ben Youssef, C. Hahn, A H. Molpeceres, C. Carrétéro, E. Jacquet, C. Deranlot, P. Bortolotti, R. Lebourgeois, J.-C. Mage, G. de Loubens, O. Klein, V. Cros, and A. Fert, Appl. Phys. Lett. **103**, 082408 (2013).

9. C. Hauser, T. Richter, N. Homonnay, C. Eisenschmidt, H. Deniz, D. Hesse, S. Ebbinghaus, G. Schmidt, and N. Weinberg, Sci. Rep. **6**, 20827 (2016).

10. P. Pirro, T. Brächer, A. V. Chumak, B. Lägel, C. Dubs, O. Surzhenko, P. Görnert, B. Leven, and B. Hillebrands, Appl. Phys. Lett. **104**, 012402 (2014).

11. H. Yu, O. d'Allivy Kelly, V. Cros, R. Bernard, P. Bortolotti, A. Anane, F. Brandl, R. Huber, I. Stasinopoulos, and D. Grundler, Sci. Rep. **4**, 6848 (2014).

12. M. B. Jungfleisch, W. Zhang, W. Jiang, H. Chang, J. Sklenar, S. M. Wu, J. E. Pearson, A. Bhattacharya, J. B. Ketterson, M. Wu, and A. Hoffmann, J. Appl. Phys. **117**, 17D128 (2015).

13. H. Yu, O. d' Allivy Kelly, V. Cros, R. Bernard, P. Bortolotti, A. Anane, F. Brandl, F. Heimbach, and D. Grundler, Nat. Commun. **7**, 11255 (2016).

14. M. Collet, O. Gladii, M. Evelt, V. Bessonov, L. Soumah, P. Bortolotti, S. O. Demokritov, Y. Henry, V. Cros, M. Bailleul, V. E. Demidov, and A. Anane, Appl. Phys. Lett. **110**, 092408 (2017).

15. V. E. Demidov, S. Urazhdin, R. Liu, B. Divinskiy, A. Telegin, and S. O. Demokritov, Nat. Commun. **7**, 10446 (2016).

16. M. Evelt, V. E. Demidov, V. Bessonov, S. O. Demokritov, J. L. Prieto, M. Muñoz, J. Ben Youssef, V. V. Naletov, G. de Loubens, O. Klein, M. Collet, K. Garcia-Hernandez, P. Bortolotti, V. Cros and A. Anane, Appl. Phys. Lett. **108**, 172406 (2016).

17. M. Collet, X. de Milly, O. d'Allivy Kelly, V.V. Naletov, R. Bernard, P. Bortolotti, J. Ben Youssef, V.E. Demidov, S.O. Demokritov, J.L. Prieto, M. Munoz, V. Cros, A. Anane, G. de Loubens, and O. Klein, Nat. Commun. **7**, 10377 (2016).

18. V. E. Demidov, S. Urazhdin, G. de Loubens, O. Klein, V. Cros, A. Anane, S. O. Demokritov, Phys. Rep. **673**, 1–31 (2017).

19. J. D. Joannopoulos, S. G. Johnson, J. N. Winn, R. D. Meade, *Photonic Crystals - Molding the Flow of Light* (Princeton University Press, Princeton, 2008).

20. G. Gubbiotti, S. Tacchi, M. Madami, G. Carlotti, A. O. Adeyeye, and M. Kostylev, J. Phys. D: Appl. Phys. **43**, 264003 (2010).

21. M. Krawczyk and D. Grundler, J. Phys.: Condens. Matter **26**, 123202 (2014).

22. S. A. Nikitov, D. V. Kalyabin, I. V. Lisenkov, A. Slavin, Yu. N. Barabanenkov, S. A. Osokin, A. V. Sadovnikov, E. N. Beginin, M. A. Morozova, Yu. A. Filimonov, Yu. V. Khivintsev, S. L. Vysotsky, V. K. Sakharov, and E. S. Pavlov, Phys.-Usp. **58**, 1002 (2015).

23. S. Tacchi, G. Gubbiotti, M. Madami, and G. Carlotti, J. Phys.: Condens. Matter **29**, 073001 (2017).

24. A. V. Chumak, A. A. Serga, and B. Hillebrands, Magnonic crystals for data processing, arXiv:1702.06701v1.

25. K. S. Lee, D. S. Han, and S. K. Kim, Phys. Rev. Lett. **102**, 127202 (2009).

26. A. V. Chumak, P. Pirro, A. A. Serga, M. P. Kostylev, R. L. Stamps, H. Schultheiss, K. Vogt, S. J. Hermsdoerfer, B. Laegel, P. A. Beck, and B. Hillebrands, Appl. Phys. Lett. **95**, 262508 (2009).

27. A. Vansteenkiste, J. Leliaert, M. Dvornik, M. Helsen, F. Garcia-Sanchez, and B. Van Waeyenberge, AIP Advances **4**, 107133 (2014).

28. V. E. Demidov, S. O. Demokritov, K. Rott, P. Krzysteczko, and G. Reiss, Phys. Rev. B **77**, 064406 (2008).

29. V. E. Demidov, M. P. Kostylev, K. Rott, P. Krzysteczko, G. Reiss, and S. O. Demokritov, Appl. Phys. Lett. **95**, 112509 (2009).

30. A. A. Serga, T. Schneider, B. Hillebrands, S. O. Demokritov, and M. P. Kostylev, Appl. Phys. Lett. **89**, 063506 (2006).

31. V. E. Demidov, S. Urazhdin, and S. O. Demokritov, Appl. Phys. Lett. **95**, 262509 (2009).

32. P. Kolodin and B. Hillebrands, J. Magn. Magn. Mater. **161**, 199 (1996).

33. A. B. Ustinov, B. A. Kalinikos, V. E. Demidov, S. O. Demokritov, Phys. Rev. B **81**, 180406(R) (2010).

34. S. O. Demokritov, A. A. Serga, A. Andre, V. E. Demidov, M. P. Kostylev, B. Hillebrands, A. N. Slavin, Phys. Rev. Lett. **93**, 047201 (2004).